\documentclass[twocolumn,showpacs,showkeywords,prb]{revtex4}
\usepackage{graphicx}
\usepackage{dcolumn}
\usepackage{amsmath}
\usepackage{latexsym}

\def\AP{{\it Ann. Phys.} }

\def\NAT{{\it Nature} }

\def\PL{{\it Phys. Lett.} }
\def\PR{{\it Phys. Rev.} }

\begin{document}
\title
{Galilean symmetry in noncommutative Gravitational Quantum Well }
\author{Anirban Saha}
\affiliation{Department of Physics and Astrophysics, \\ West Bengal State University, Barasat,\\ West Bengal, India.}

\date{\today}


\begin{abstract}
{A thorough analysis of Galilean symmetries for the gravitational well problem on a noncommutative plane is presented. A complete closure of the one-parameter centrally extended Galilean algebra is realised for the model. This implies that the field theoretic model constructed to describe noncommutative gravitational quantum well in \cite{ani} is indeed independent of the coordinate choice. Hence the energy spectrum predicted by the model can be associated with the experimental results to establish the upper-bound on time-space noncommutative parameter.  Interestingly, noncommutativity is shown to increase the gravitational pull on the neutron trapped in the gravitational well.}
\end{abstract}


\maketitle
\section{Introduction}
The idea of noncommutative (NC) space time \cite{sny} where the coordinates $x^{\mu}$ satisfy the
noncommutative algebra
\begin{equation}
\left[x^{\mu}, x^{\nu}\right] = i \Theta^{\mu \nu}
\label{ncgometry}
\end{equation}
has gained prominence in the recent past \cite{sw} and field theories defined over this NC space are currently a subject of very intense research \cite{sz}. A wide range of theories are being formally studied in a NC perspective encompassing various gauge theories \cite{rbgauge} including gravity \cite{grav}.
Apart from studying the formal aspects of the NC geometry certain possible phenomenological consequences have also been investigated \cite{jabbari1, jabbari2, jabbari3, cs, rs1, rs2, rs3, rs4, rs5, rs6, rs7, rs8, rs9, rs10}. Often such NC theories produce results which are deformed from their commutative counterpart by the presence of NC correction terms. These correction terms are usually proportional to various orders of the NC parameters \cite{sch, sch1, RN, chichian}. Naturally a part of the endeavor is spent in finding the order of the NC parameter and in exploring its connection with observations \cite{cst, mpr, carol}.  

A particular piece of the scenario is the quantum well problem which has emerged in recent GRANIT experiments by Nesvizhevsky {\it et al.} \cite{nes1, nes2, nes3} who detected the quantum states of the neutrons trapped in earth's gravitational field. Their experimentally determined energy spectrum has been compared with the theoretically predicted energy spectrum of a quantum mechanical model considered by Bertolami {\it et al.} \cite{bert0, bert1} and Banerjee {\it et al.} \cite{RB} to set an upper bound on the momentum space NC parameters. Naturally, noncommutativity is introduced among the phase space variables and noncommutativity in the time-space sector was ignored since in QM as such space and time could not be treated on an equal footing. 

To introduce time-space noncommutativity a second quantized theory was proposed in \cite{ani} where the NC quantum well problem is pictured as a non-relativistic field theory
coupled with an external gravitational field. This work resulted in a upper-bound estimation for the time-space NC parameter, consistent with the existing results of \cite{bert0, bert1} and \cite{RB}. More recently, another work \cite{new} have claimed to impose an upper-bound on the space-space NC parameter by using the GRANIT experimental data.

This recent trend of associating the predictions of NC theoretical models of the gravitational well with experimental results is very encouraging. However, it is perhaps appropriate to pause before any further development in this topic to assess to what extent such associations make physical sense. For them to be physically meaningful, these NC theories have to be independent of the coordinate choice. Hence, a crucial underlying assumption in all the above-mentioned works has been the coordinate independence of the quantum well system constructed on the NC plane. In other words the system has to be invariant under the Galilean transformation. This is also a crucial requirement in connection with the treatment done in \cite{ani} since single particle quantum mechanics can be viewed as the one-particle sector of quantum field theory for Galilean-invariant systems only \cite{Nair, RB1, bcsgas}.

That the NC algebra (\ref{ncgometry}) violates the active Lorentz symmetry is manifest owing to the constancy of the NC tensor $\Theta^{\mu \nu}$ in (\ref{ncgometry}) . However, the same can not be assured for the Galilean symmetries in general. For example, the boost symmetry is violated in NC planer system \cite{bcsgas} where time-space noncommutativity is ignored, although the algebra (\ref{ncgometry}) is preserved with $\Theta^{0i} = 0$ under Galileo boost. The situation is further complicated by the presence of a non-zero $\Theta^{0i}$ in \cite{ani}. It is well-known that space rotations are not automorphisms of the algebra (\ref{ncgometry}) and therefore it is not invariant under $O(2)$ rotations for $\Theta^{0i} \neq 0$. However, when there are spectral degeneracies of a time-independent Hamiltonian on a commutative space-time which are due to symmetries, they persist for $\Theta^{0i} \neq 0$ \cite{bala0, bala3}. Therefore it is all the more important to examine the algebra of the generators of the theory for any possible violation or otherwise of Galilean symmetries. Hence, in this paper we like to address this issue from the same field theoretic approach presented in \cite{ani}. 

Our NC field theoretic modeling of the gravitational quantum well with noncommutativity among time and space coordinates, i.e., a non-zero $\Theta^{0i}$ gains further perspective in connection with yet another non-triviality of NC field theories, namely, the contentious unitarity issue. There are claims that introduction of space-time noncommutativity spoils unitarity \cite{gomis, gaume, sei}. Specifically, in \cite{gomis, sei} it was argued that because of the presence of infinite time derivatives, space-time noncommutative theories can not be quantised properly. In contrast, in a series of fundamental papers, Doplicher {\it et al.} \cite{bala, bala1} have studied (\ref{ncgometry}) in complete generality, without assuming $\Theta^{0i} = 0$ and developed unitary quantum field theory's which are ultraviolet finite to all orders. Based on them in \cite{bala0, bala3} a systematic development of unitary quantum mechanics followed. Ho {\it et al.} in \cite{ho1} discussed a perturbative approach to higher-derivative theories (where the Lagrangean contains higher/infinite order time derivatives)  following \cite{wood} where they have constructed a consistent Poission structure and Hamiltonian. They alos gave a formal proof that the process can be carried out to all order. Further, in \cite{liao} it was shown that perturbative unitarity can be successfully maintained if one takes care of the explicit Hermiticity of the Lagrangean. Even a canonical formalism can be developed by introducing an additional space-time dimension \cite{kamimura}. Following  \cite{dop1} where it was shown that unitarity problem is not inherent, but is due to ill-defined time-ordered product, Rim {\it et al.} in \cite{rim1, rim2} have demonstrated how perturbative analysis in the space-time NC field theories respect the unitarity if the S-matrix is defined with proper time-ordering and free spectral function is used instead of Feynmann propagetor.  In \cite{sch, sch1}, it was demonstrated in the context of (1 + 1)-dimensional NC Schwinger model, where noncommutativity among the time and space coordinate is essential, that a straight-forward perturbative approach retaining terms up to first order in the NC parameter leads to a unitary S-matrix and also ensures causality. In the present paper we analyse the space-time symmetry of a similar model where noncommutativity among the time and space coordinates takes a crucial role since it modifies the energy spectrum. In fact, the model is such that the space-space NC parameter can be scaled out of the system and non-trivial NC correction comes only from the time-space sector \cite{ani}. 

In the following section we shall briefly summaries the NC field theory describing the NC gravitational quantum well system. Apart from setting the platform for the present problem this will also help us fix the notations. We shall construct the Galilean symmetry generators from the energy-momentum tensor and compute their algebra in section 3. Due to the absence of a metric in Galilean space-time the boost generators can not be associated with the appropriate components of the total angular momentum tensor \cite{bcam, pkm}. Hence the construction of Galileo boost shall need separate attention here. The time evolution of the momentum generator exhibits that the presence of time-space noncommutativity increases the gravitational pull on the trapped neutron; an upper bound estimation of this excess pull is calculated in section 4 using the present upper bound on the time-space NC parameter $\eta$. In this connection we also discuss the different bounds on various NC parameters available in the literature and their consistency with the upper-bound on time-space NC parameter obtained by considering the gravitational well problem \cite{ani}. We conclude in section 5.

\section{The NC Schr\"{o}dinger theory}
In this section we shall briefly summaries the NC field theory of a neutron trapped in earth's gravitational field. 
We choose to work in the deformed phase space with the ordinary product replaced by the star product \cite{carol, Dayi, bcsgas, our1, our2}. In this formalism the fields are defined as functions of the phase space variables and the redefined product of two fields $\hat \phi(x)$ and $\hat \psi(x)$ is given by 
\begin{equation}
\hat \phi(x) \star \hat \psi(x) = \left(\hat \phi \star \hat \psi \right)(x) = e^{\frac{i}{2}
\theta^{\alpha\beta}\partial_{\alpha}\partial^{'}_{\beta}}
  \hat \phi (x) \hat \psi(x^{'})\big{|}_{x^{'}=x.}
\label{star}
\end{equation}
In star-product formalism the action for a NC Schr\"{o}dinger field $\hat \psi$ coupled with background gravitational field reads 
\begin{eqnarray} 
\hat S = \int dx \hspace{0.5mm} dy\hspace{0.5mm} dt \hspace{1.0mm} \hat \psi^{\dag}\star \left[i \hbar \partial_{0} + \frac{{\hbar}^{2}}{2m} \partial_{i}\partial_{i} - m g\hat{x} \right] \star \hat \psi
\label{NCaction} 
\end{eqnarray}
The above action describes a system in a vertical $xy$ ($i = 1, 2$) plane where the external gravitational field is taken parallel to the $x$-direction. 
Under $\star $ composition the Moyal bracket between the coordinates is
\begin{eqnarray} 
\left[\hat x^{\mu},\hat x^{\nu}\right]_{\star} = i\Theta^{\mu\nu} = i\left(\begin{array}{ccc}
0 & -\eta & -\eta^{\prime} \\
\eta & 0 & \theta\\
\eta^{\prime}  & -\theta & 0\\
\end{array}\right)
\label{NCpara}
\end{eqnarray}
where $\mu, \nu$ take the values $0, 1, 2$. Spatial noncommutativity is denoted by $\Theta^{12} = \theta$ and noncommutativity among time and the two spatial directions are denoted by the parameters $\Theta^{10} = \eta$ and $\Theta^{20} = \eta^{\prime}$. 

Expanding the $\star$-product and rescaling the field variables and mass\footnote{Note that the mass term has been rescaled and interpreted as the observable mass in \cite{ani}. A similar charge rescaling of NC origin in context of NC QED was shown in \cite{carol}.}
 by 
\begin{eqnarray}
\psi \mapsto \tilde{\psi} &=& \sqrt{\left(1 - \frac{\eta}{2 \hbar} m g\right)}\psi \nonumber\\
\tilde m &=& \left(1 - \frac{\eta}{2 \hbar} m g \right) m
\label{scal1}
\end{eqnarray}
we get
\begin{eqnarray} 
\hat S = \int dx dy dt \tilde{{\cal{L}}} &=& \int  dx  dy dt  \tilde{\psi}^{\dag} \left[i \hbar \partial_{t}  +  \frac{{\hbar}^{2}}{2\tilde{m}} \left(\partial_{x}{}^{2} + \partial_{y}{}^{2}\right) \right. \nonumber \\
 &&- \left.\tilde {m}g x  -  \eta \left(\frac{\tilde{m}{}^{2} g^{2}}{\hbar}\right)x \right] \tilde{\psi}
\label{Caction2} 
\end{eqnarray} 
where a first derivative term is absorbed in the $\partial_{y}{}^{2}$ by rewriting
\begin{eqnarray}
\partial_{y} = \left(\partial_{y} -  \frac{i \theta}{2 \hbar^{2}} \tilde {m}{}^{2} g \right) .
\label{scale3}
\end{eqnarray}
Note that the structure of the Lagrangean is such that expanding the star product only gives corrections to first order in the NC parameters and all the higher order terms vanish.

Using a Hamiltonian description it was shown in \cite{ani} that these rescaled variables are the proper canonical pair of fields in the sense that they satisfy the standard form of the PB relation. The only primary constraint of the theory comes from the definition of the canonical momenta corresponding to the field variable\footnote{Note that in the argument we collectively refer to both the spatial coordinates by $x$ in the reminder of the paper.} 
\begin{eqnarray}
\tilde{\pi}{}_{\psi} \left(x\right) =  i \hbar \tilde{\psi}^{\dag}\left(x\right)
\label{momenta}
\end{eqnarray}
The canonical Hamiltonian density $\tilde{{\cal{H}}}{}_{c}$ can be computed by a Legendre transformation to obtain 
\begin{eqnarray}
\tilde{{\cal{H}}}{}_{c} & = & \tilde{\pi}{}_{\psi} \dot{\tilde{\psi}} - \tilde{{\cal{L}}}  \nonumber\\
& = & - \frac{{\hbar}^{2}}{2\tilde{m}} \tilde{\psi}^{\dag}\partial_{i}{}^{2}\tilde{\psi}  + \tilde{m} g \left(1 + \eta \frac{\tilde{m} g}{\hbar}\right)\tilde{\psi}^{\dag} x \tilde{\psi}
\label{H}
\end{eqnarray}
Note that the Hamiltonian is real and therefore, from a first quantized point of view, we can fairly say that the theory respects unitarity.
From eqn.(\ref{momenta}) clearly, the system is inflicted with the second class constraints. We exploit the fact that the Lagrangean is first order in the time derivative and apply the Faddeev--Jackiw (F--J) scheme to read off the basic brackets:
\begin{eqnarray}
\left\{\tilde \psi(x), \tilde \psi^{\dag}\left( x^{\prime}\right)\right\} = - \frac{i}{\hbar} \delta^{2}\left(x -  x^{\prime}\right)
\label{commutation}
\end{eqnarray}
Since in the F--J approach the second class constraints have automatically been taken in to account, in the reminder of the paper we impose the relation (\ref{momenta}) strongly. Obviously, with (\ref{commutation}) the canonical Hamiltonian generates the correct time-evolution of the field variables $\tilde{\psi}(x)$ 
\begin{eqnarray} 
\left[i \hbar \partial_{t}  +  \frac{{\hbar}^{2}}{2\tilde{m}} \left(\partial_{x}{}^{2} + \partial_{y}{}^{2}\right) - \tilde {m}g x  -  \eta \left(\frac{\tilde{m}{}^{2} g^{2}}{\hbar}\right) x \right] \tilde{\psi} = 0 \nonumber\\
\label{eqm} 
\end{eqnarray} 
Having described the basic structure of the theory we are now in a position to analyse the space-time symmetries it admits. In the next section we shall take up this task.
\section{The Galilean generators and their algebra}
In this section we shall use the Noether's theorem to identify the Galilean generators for the gravitational well system. We shall subsequently compute their algebra at the classical level to check for any possible violation or otherwise of the Galilean symmetry. We shall first consider the spatial translation generator and rotation generator and their time evolution. The generator of the Galileo boost will be taken up next. Unlike the Lorentz boost, Galileo boost can not be thought of as some kind of rotation in space time since Galilean space time is not endowed with a metric. So, contrary to the relativistic case, the boost can not be interpreted as a component of the total angular momentum tensor. Hence we shall compute the boost generator from the first principle. Once we have the expression for the generators of all type of Galilean transformations we shall compute their algebra using the fundamental F--J brackets (F--JB) (\ref{commutation}) of the theory and check if they give rise to a closure of the Galilean algebra.
\subsection{Spatial translation and the momentum generator}
We begin with the generator of spatial translation, 
\begin{eqnarray}
P_{i} = \int d^{2}x T_{0i}\left( x \right) 
\label{MG}
\end{eqnarray}
where $T_{0i}$ is the momentum density given by 																	 
\begin{eqnarray}
T_{0i} = \frac{\partial \tilde{{\cal{L}}}}{\partial \dot{\tilde{\psi}}}\partial_{i}\tilde{\psi} - \delta_{0i}\tilde{{\cal{L}}} =  \tilde{\pi}{}_{\psi}\partial_{i} \tilde{\psi} = i\hbar \tilde{\psi}{}^{\dag}\partial_{i} \tilde{\psi}
\label{EMT}
\end{eqnarray}
Note that we have replaced the conjugate momenta using (\ref{momenta}) since they hold strongly now. It can be easily checked using the fundamental F--JB relations (\ref{commutation}) that $P_{i}$ generate the proper spatial translation .
\begin{eqnarray}
\left\{\tilde{\psi}\left(x \right), P_{i }\right\} &=&  \partial_{i}\tilde{\psi}\left( x \right) \nonumber\\
\left\{\tilde{\psi}{}^{\dag}\left(x \right), P_{i }\right\} &=&  \partial_{i}\tilde{\psi}{}^{\dag}\left( x \right)
\label{ST}
\end{eqnarray}
The generator of time evolution is given by 
\begin{eqnarray}
H &=& \int d^{2}x T_{00}\left(x \right) = \int d^{2}x \left[\frac{\partial \tilde{{\cal{L}}}}{\partial \dot{\tilde{\psi}}}\dot{\tilde{\psi}} - \delta_{00}\tilde{{\cal{L}}} \right] \nonumber \\
&=& \int d^{2}x \left[- \frac{{\hbar}^{2}}{2\tilde{m}} \tilde{\psi}{}^{\dagger}\partial_{i}{}^{2} \tilde{\psi} 
+ \tilde{m}g \left(1 +\frac{\eta \tilde{m}g}{\hbar}\right)\tilde{\psi}{}^{\dagger} x \tilde{\psi}\right]\nonumber\\
\label{T00}
\end{eqnarray}
which of course corresponds to the canonical Hamiltonian density ${\cal{H}}_{c}$ already derived in (\ref{H}). The time evolution of momentum generator gives
\begin{eqnarray}
\left\{P_{i}, H\right\} = \tilde{m}g \left(1 +\frac{\eta \tilde{m}g}{\hbar}\right)\int d^{2}x   
\tilde{\psi}{}^{\dagger} \partial_{i}x  \tilde{\psi}
\label{PH}
\end{eqnarray}
For $i = 2$, $P_{2} = P_{y}$, the usual conservation of momentum in $y$-direction follows. However, for $i = 1$, $P_{1} = P_{x}$, we get the time evolution of the momentum generator in the direction of the gravitational interaction, this naturally gives the gravitational force acting on the trapped neutron
\begin{eqnarray}
\left\{P_{x}, H\right\} &=& \tilde{m}g \left(1 +\frac{\eta \tilde{m}g}{h}\right)\int dx dy    
\tilde{\psi}{}^{\dagger} \partial_{x}x  \tilde{\psi}\nonumber\\
&& =  \tilde{M}g \left(1 +\frac{\eta \tilde{m}g}{\hbar}\right) 
\label{PxH}
\end{eqnarray}
where 
\begin{eqnarray}
\tilde{M} = \tilde{m}\int dx dy \tilde{\psi}{}^{\dagger} \tilde{\psi}
\label{m}
\end{eqnarray}
is the total mass. Interestingly, (\ref{PxH}) tells us that the presence of time-space noncommutativity increases the gravitational pull on the neutron by a factor of $\frac{\eta \tilde{m}g}{\hbar}$. This is a significant physical effect of time-space noncommutativity and we shall shortly estimate the maximum possible increase of the gravitational pull using the upper bound estimate established in \cite{ani}.

It can be shown that under the phase transformation $\tilde{\psi} \to e^{i m \phi} \tilde{\psi}$ the Lagrangean remains invariant for infinitesimal $\phi$ and $\tilde{M}$ in (\ref{m}) is the generator of this transformation \cite{Hagen}.  This is the first central extension of the Galilei group by a one-dimensional Abelian group where $\tilde{M}$ commutes with all the  operators of the group. 

\subsection{$SO\left(2\right)$ Rotation and angular momentum generator}
Proceeding similarly, the angular momentum $J$ is given as
\begin{eqnarray}
J = \int d^{2}x \epsilon_{ij}x_{i}T_{0j}
\label{J }
\end{eqnarray}
Using (\ref{momenta}) and (\ref{EMT}) this expression is simplified to 
\begin{eqnarray}
J =  i\hbar \int d^{2}x \epsilon_{ij} x_{i}\tilde{\psi}{}^{\dag}\partial_{j} \tilde{\psi}
\label{J1 }
\end{eqnarray}
which generates appropriate  $SO\left(2\right)$ rotation of the fields.
\begin{eqnarray}
\left\{\tilde{\psi}\left(x \right), J \right\} =  \epsilon_{ij} x_{i} \partial_{j}\tilde{\psi} 
\label{rotation}
\end{eqnarray}
Note that $J$ consists of only the orbital part of the angular momentum as we have ignored the spin degrees of freedom for the field $\tilde{\psi}$, so that it transforms as an $SO\left(2\right)$ scalar.

Using (\ref{commutation}) it can be easily checked that the F--JB among the rotation generator $J$ and time evolution generator $H$ gives 
\begin{eqnarray}
\left\{J, H \right\} = -\tilde{m}g \left(1 +\frac{\eta \tilde{m}g}{\hbar}\right) \int dx dy \tilde{\psi}{}^{\dagger} y  \tilde{\psi}
\label{JH}
\end{eqnarray}
A non-zero r.h.s. in (\ref{JH}) apparently implies that the angular momentum generator is not conserved and the system does not support rotational symmetry. This, however, is not the case since the system is symmetric in $y$-direction and the fields $\tilde{\psi}$ and $\tilde{\psi}{}^{\dagger}$ must have a certain parity under the transformation $y \to -y$. So the integration appearing in (\ref{JH}) vanishes and the conservation of angular momentum prevails.
\subsection{The boost generator}
So far we have derived all the generators necessary to construct the two-dimensional (Euclidean) $E\left(2\right)$ algebra. However, the construction of the full Galilean algebra requires another set of generators corresponding to the two boost transformations in the two spatial directions. As has been mentioned earlier, unlike the relativistic case, the boost here is not a part of the total angular momentum. So we shall derive them by analysing the system from the first principle following \cite{bcam}. 

Let us consider an infinitesimal Galileo boost in the $x$-direction:
\begin{eqnarray}
t \to t^{\prime} &=& t, \qquad x \to x^{\prime} = x - vt \nonumber\\
y \to y^{\prime} &=& y
\label{IGB}
\end{eqnarray}
where the velocity $v$ is infinitesimal. The canonical basis corresponding to the unprimed and primed variables are given by 
$\left( \partial/ \partial t,  \partial/ \partial x_{i} \right)$ and $\left( \partial/ \partial t^{\prime},  \partial/ \partial x_{i}{}^{\prime} \right)$, respectively. They are related by 
\begin{eqnarray}
\frac{\partial}{\partial t^{\prime}} &=& \frac{\partial}{\partial t} + v \frac{\partial}{\partial x} \nonumber \\
\frac{\partial}{\partial x_{i}{}^{\prime}} &=& \frac{\partial}{\partial x_{i}}
\label{GBT}
\end{eqnarray}

Since, in the first quantized version of the theory $\tilde{\psi}$ is interpreted as probability amplitude \cite{ani}, it is expected that the probability density $\tilde{\psi}{}^{\dagger}\tilde{\psi}$ will remain invariant under the Galileo boost (\ref{IGB}), i.e. $ \tilde{\psi}{}^{\dagger}\left( x, t \right)\tilde{\psi}\left( x, t \right) = \tilde{\psi}{}^{\dagger}\left( x^{\prime}, t^{\prime} \right)\tilde{\psi}\left( x^{\prime}, t^{\prime} \right) $. Thus we expect $\tilde{\psi}$ to change at most by a phase under (\ref{IGB}). Hence we make the following ansatz \cite{bcsgas}:
\begin{eqnarray}
\tilde{\psi}\left( x, t \right) \to \tilde{\psi}\left( x^{\prime}, t^{\prime} \right) &=& e^{i v \xi \left( x, t \right)} \tilde{\psi} \left( x, t \right) \nonumber \\
&&  \approx  \left[1 + i v \xi \left( x, t \right) \right] \tilde{\psi}\left( x, t \right)
\label{ansatz}
\end{eqnarray}
Now we demand the covariance of the equation of motion, i.e. (\ref{eqm}) retains its form in both unprimed and primed coordinates. Using (\ref{IGB}) and (\ref{GBT}) in (\ref{eqm}) we see that this requirement leads to the following coupled differential equation of the boost parameter $\xi$.
\begin{eqnarray}
\frac{i {\hbar}^{2}}{2 \tilde{m}}\partial_{x}{}^{2} \xi - \hbar \partial_{t} \xi + \tilde{m}g \left(1 +\frac{\eta \tilde{m}g}{\hbar}\right)t = 0; \nonumber\\
\frac{\hbar}{\tilde{m}} \partial_{x} \xi  + 1 = 0
\label{DE}
\end{eqnarray}
Solving (\ref{DE}) we get 
\begin{eqnarray}
\xi = \frac{\tilde{m}}{\hbar} x + \frac{1}{2 \hbar}\tilde{m}g \left(1 +\frac{\eta \tilde{m}g}{\hbar}\right) t^{2}
\label{Sol}
\end{eqnarray}
Since the boost parameter admits real solution, the wave function preserves its norm under (\ref{ansatz}) and we conclude that boost transformation in the direction of the gravitational field is a symmetry in the NC gravitational well system. This is similar to the situation encountered in \cite{bcsgas} where a boost in the direction of the electric field was found to be a symmetry of the system. 
The functional change of the field under this transformation is obtained as 
\begin{eqnarray}
\delta_{0}\tilde{\psi}\left( x, t \right) &=& v \left[t \partial_{x}\tilde{\psi}\left( x, t \right) - \frac{i \tilde{m}}{\hbar} x \tilde{\psi}\left( x, t \right)\right.  \nonumber \\
&& + \left. \frac{i}{2 \hbar}\tilde{m}g \left(1 +\frac{\eta \tilde{m}g}{\hbar}\right) t^{2}\tilde{\psi}\left( x, t \right)\right]
\label{FC}
\end{eqnarray}
Expression for the boost generator $K_{x}$ can be read off by comparing the functional change of the field $\tilde{\psi}$ (\ref{FC}) with the result of the F--JB among the field $\tilde{\psi}$ and the boost generator multiplied by the velocity $v$ :
\begin{eqnarray}
v\left\{\tilde{\psi}\left( x \right), K_{x} \right\} = \delta_{0}\tilde{\psi}\left( x \right)
\label{Kpsi}
\end{eqnarray}
which gives                                                                         
\begin{eqnarray}
K_{x} &=& t P_{x} + \tilde{m} \int dx^{\prime}dy^{\prime}  \tilde{\psi}{}^{\dagger}\left( x^{\prime} \right)  x^{\prime} \tilde{\psi}\left( x^{\prime} \right) \nonumber \\
&& - \frac{1}{2}g \left(1 + \frac{\eta \tilde{m} g}{\hbar}\right) t^{2} \tilde{M} 
\label{Kx}
\end{eqnarray}

Following the same scheme we can construct the Galileo boost generator in the direction perpendicular to the gravitational field (\i.e. $K_{y}$) which gives
\begin{eqnarray}
K_{y} &=& t P_{y} + \tilde{m} \int dx^{\prime}dy^{\prime} y^{\prime} \tilde{\psi}{}^{\dagger}\left( x^{\prime} \right) \tilde{\psi}\left( x^{\prime} \right)
\label{Ky}
\end{eqnarray}
This result differs from the result of \cite{bcsgas} where, in the context of a NC Schr\"{o}dinger field theory coupled to external gauge fields, the boost transformation perpendicular to the external (electric) field led to a non-unitarity of the wave function signifying a violation of the boost symmetry. Our present analysis of the gravitational well system shows that boost transformations, both in the direction parallel and perpendicular to the external (gravitational) field, lead to unitary transformations of the field variable. Thus boost along as well as perpendicular to the external field are consistent symmetries here.
\subsection{The NC Galilean Algebra}
What remains is to compute the algebra among various Galilean generators. The F--JBs among the spatial translation and rotation generators along with the time evolution generator form the closed $E\left(2\right)$ algebra:
\begin{eqnarray}
\left\{P_{x}, P_{y}\right\} &=& 0 \nonumber\\
\left\{P_{x}, H\right\} &=& \tilde{M}g \left(1 +\frac{\eta \tilde{m}g}{\hbar}\right) \nonumber\\
\left\{P_{y}, H\right\} &=&0 \nonumber\\
\left\{J, H\right\} &=& 0 \nonumber\\
\left\{P_{i}, J\right\} &=& \epsilon_{ij}P_{j}
\label{E2}
\end{eqnarray}
The presence of a non-zero right hand side in the second equation merely signifies the force acting on the trapped neutron. However, this can easily be related with the one-parameter central extension associated with the total mass of the system. Also note that the NC parameter $\eta$ entaring the algebra is not just a consequence of coordiante choice but a result of writing the theory in terms of proper canonical variables. To complete the Galilean algebra we include the boost sector at this point.  
The boost generators give vanishing F--JBs among themselves
\begin{eqnarray}
\left\{K_{x}, K_{y} \right\} &=& 0
\label{KxKy}
\end{eqnarray}
Their F--JB with the remaining Galilean generators are now worked out which give 
\begin{eqnarray}
\left\{P_{i}, K_{j} \right\} &=& \delta_{ij} \tilde{M}\nonumber \\
\left\{K_{i}, J \right\} &=& \epsilon_{ij}K_{j} \nonumber \\
\left\{K_{x}, H \right\} &=& -\left( P_{x} -t \tilde{M} g \right)\nonumber\\
\left\{K_{y}, H \right\} &=&-P_{y} 
\label{PiKj}
\end{eqnarray}
Once again, we note that since $\tilde{M}$ is also a generator, its appearance in the third relation does not violate the closure of the Galilean algebra.
The first central extension also gives vanishing F--JBs with all the generators.
\begin{eqnarray}
\left\{P_{i}, \tilde{M}\right\}&=& 0 \nonumber\\
\left\{J, \tilde{M}\right\}&=& 0 \nonumber\\
\left\{H, \tilde{M}\right\}&=& 0 \nonumber\\
\left\{K_{i}, \tilde{M}\right\}&=& 0
\label{CE}
\end{eqnarray}
This concludes our computation of the Galilean algebra. As is manifest from the above equations (\ref{E2}, \ref{KxKy}, \ref{PiKj}, \ref{CE}) the generators form a closure and hence Galilean symmetry is preserved for the  gravitational well system.
\section{Upper-bound on the time-space NC parameter and its effect}
In section 3.1, the time-evolution of the momentum generator in the direction of the external gravitational field was computed in equation (\ref{PxH}) and naturally, this produces the gravitational force acting on the trapped neutron. Interestingly, along with the expected force term ${\tilde{M}}g$ a extra piece appears which is proportional to the time-space NC parameter $\eta$. This clearly revels that time-space noncommutativity increases the gravitational pull on the trapped neutron. Therefore using the upper-bound of $\eta$ estimated in our earlier work\cite{ani}, we can estimate to what extent the time-space NC parameter enhances the gravitational pull. This may indeed present a simplistic scenario where we can detect the noncommutativity of space-time. Before going in to that we shall briefly summarise the existing upper-bounds on various NC parameters.In perticular, we shall demonstrate the consistancy of the bound on the time-space NC parameter obtained by considering the gravitational well system in \cite{ani} with the other bounds existing in the literature.
\subsection{The existing upperbounds on various NC parameters}
As noncommutativity is motivated from string theory and quantum gravity, its effect is expected to show up at the Plank scale. Nevertheless, experimentally accessible scales should be explored, especially since the current research on large extra dimensions can potentially bring down the four-dimensional Plank scale. Therefore a considerable amount of work has been done to work out bounds on the NC parameters. Verious authors have worked out different bounds on the space-space NC parameter which ranges from $\theta \lesssim \left(10 {\rm TeV}\right)^{-2}$ \cite{carol} to $\theta \lesssim \left(30 {\rm MeV}\right)^{-2}$ \cite{stern}. In \cite{pmh} it was argued that NC parameters for different particles should be different, specifically particles with opposite charge should bear opposite noncommutativity, which makes their relative coordinate commutating. In the Hydrogen atom problem corrections then resulted in the Lamb shifts due to noncommutativity of just the electrons in \cite{ch}. It was further argued there that since QED effects may dominate over noncommutativity the nucleus should be treated as a commutative object. However, in absence of a fully understood theory of NC QED, in \cite{stern} the authors has considered the nucleus of the Hydrogen atom to be a NC point charge and worked out the most recent bound on the space-space NC parameter. There the $\theta \lesssim \left(6 {\rm GeV}\right)^{-2}$ bound was found for the test charge i.e. electron noncommutativity whereas a much lower bound was found for the proton noncommutativity $\theta \lesssim \left(30 {\rm MeV}\right)^{-2}$.

Similarly, also for momentum NC parameters, several works have been put forward \cite{bert0, RB} where the upper bound on space-space NC parameter found by Carroll {\it et. al} \cite{carol} has to be used. Surprisingly, not much attention has been paid to the time-space NC parameter, owing to the ambiguity concerning the uniterity issue. However, in the introduction we have provided arguments in favour of a perturbative treatement of theories which incorporates time-space noncommutativity. In our earlier work we have estimated the bounds on time-space NC parameter by formulating a NC field theory of gravitational quantum well and comparing our theoretical energy spectrum with experimental data \cite{nes1, nes2, nes3}. In fact, that such an association of the experimental data with our model can be made, is established in the present paper by showing that our NC model of gravitational quantum well respects the Galilean symmetry at a field theoretic level. Presently we give a rough comparision of the bound we found in \cite{ani} with the existing results in the literature \cite{bert0, RB}. 

In \cite{bert0} the upper bound on the fundamental momentum scale was calculated to be 
\begin{eqnarray}
\Delta p & \lesssim & 4.82 \times 10^{-31}\ \mathrm{kg\ m\ s^{-1}}
\label{bert_scale1}
\end{eqnarray}
Since $E \approx \frac{p_{y}{}^{2}}{2\tilde{m}}$ so 
\begin{eqnarray}
\Delta E \approx \frac{p_{y}}{\tilde{m}}\ \Delta p_{y} = v_{y} \Delta p_{y} \lesssim 31.33 \times 10^{-31} kg\ m^{2}\ s^{-2}
\label{bert_scale2}
\end{eqnarray}
Here we have used the value of $v_{y} = 6.5\ \mathrm{m\ s^{-1}}$ used by the GRANIT experiment group. Using this value of $\Delta E$ in the time energy uncertainty relation $\Delta E \Delta t \geq \hbar$, we find 
\begin{eqnarray}
\Delta t \geq \frac{\hbar }{\Delta E} = 3.38 \times 10^{-4}\ s
\label{bert_scale3}
\end{eqnarray}
Hence uncertainty in time-space sector can be calculated using the results of \cite{bert0} as
\begin{eqnarray}
\Delta x \ \Delta t \sim  3.38 \times 10^{-18}\ m\ s
\label{bert_scale4}
\end{eqnarray}
where we have taken $\Delta x = 10^{-15} \ \mathrm{m}$. Note that this is the value used in \cite{bert0}. On the other hand in \cite{ani} we have worked out the upperbound on the parameter $\eta$ as 
\begin{eqnarray}
\eta = - i \ \left[x^{1}, x^{0}\right] & \lesssim & 2.843\times 10^{-9}\ \mathrm{m \ s}
\label{bert_scale5}
\end{eqnarray}
Restoring the $c$-factor back in (\ref{bert_scale5}) we get
\begin{eqnarray}
 - i \ \left[x , t \right] = \frac{\eta}{c} = & \lesssim & 9.51 
 \times 10^{-18} \mathrm{m \ s}
\label{bert_scale6}
\end{eqnarray}
Using the variance theorem \cite{jjs} for the commutation relation in (\ref{bert_scale6}) we can write 
\begin{eqnarray}
 \Delta x \ \Delta t \geq \frac{1}{2}\frac{\eta}{c} \sim 4.75 \times 10^{-18} \mathrm{m \ s}
\label{bert_scale7}
\end{eqnarray}
Clearly, the value of the upper bound on $\eta$ turned out to be consistent with the results of \cite{bert0, bert1, RB}. 

Note that although in \cite{ani} we have constructed our model on a field theoretic platform, this was done only to bring out the role of time-space noncommutativity in our model, something which could not be done from a quantum mechanical starting point since time and space does not share equal status in quantum mechanics. Once we obtained the perturbative correction term in the corresponding commutative equivalent field theory we switched back to the first quantised picture and did the quantum mechanical analysis. Since first and second quantized formalisms are equivalent as far as Galilean systems are concerned the upper-bound on the time-space NC parameter found in \cite{ani} can be viewed as a quantum mechanical result. 

 Interestingly, in a very recent paper \cite{match} another upper-bound has been found on the time-space NC parameter by considering the Hydrogen atom spectrum which resembles quite closely with our result in \cite{ani}. Considering the simplistic treatement done in both the cases \cite{ani, match} it is quite remarkable that two independent and unrelated phenomenological considerations, namely the trapping of cold neutron in a gravitational well and the study of Hydrogen atom spectrum, should give similar bounds on the parameter. 

\subsection{Upper-bound estimation of the excess pull on the trapped neutron}
We shall now estimate the excess pull on the trapped neutron. Using (\ref{PxH}) it can be written as  
\begin{eqnarray}
\Delta F = \left(\frac{\eta \tilde{m}g^{2}}{\hbar}\right) \tilde{m}\int dx dy \tilde{\psi}{}^{\dagger}\tilde{\psi} 
\label{EF}
\end{eqnarray}
Note that the integration along with the factor $\tilde{m}$ is interpreted as the mass generator in (\ref{m}). But it can also be interpreted, without $\tilde{m}$, as the conserved total probability of the particle, and set to unity.
\begin{eqnarray}
\int dx dy \tilde{\psi}{}^{\dagger}\tilde{\psi}  = 1
\label{unity}
\end{eqnarray}
We calculate the excess pull using the following values of the constants appearing in the expression\footnote{Note that while using the upper-bound of $\eta$ we restore the c-factor and actually use the upper-bound of $\frac{\eta}{c}$ since we are using SI unit system.} (\ref{EF})
\begin{eqnarray}
\hbar &=& 10.59 \times 10^{-35} {\rm {Js}} \nonumber \\
g &=& 9.81 {\rm{m s}}{}^{2} \nonumber\\
\tilde{m} &=& 167.32 \time 10^{-29} \nonumber\\
\frac{\eta}{c}&=& 9.51 \time 10^{-18} {\rm{m s}} 
\label{constants}
\end{eqnarray}
which gives 
\begin{eqnarray}
\delta F 
\leq 2.42 \times 10^{-35} {\rm{N}}
\label{EF}
\end{eqnarray}
Note that, similar to \cite{ani} we have restored the $c$-factor with $\eta$ since we are computing in S.I units.
The commutative part of the force is approximately $1.64 \times 10^{-26} {\rm{N}}$. So time-space noncommutativity can increase the gravitational force on the neutron at most by $0.147 \times 10^{-6}\%$. Note that ratio $\frac{\Delta F}{F} = \frac{\eta \tilde{m} g}{\hbar}$ depends on the mass of the trapped particle, thus heavier the particle trapped in the well, bigger will be the NC correction to the force acting on it.

\section{Conclusions}
We have analysed the space time symmetries of a non-relativistic system on a two-dimensional noncommutative (NC) plane. The system contains a particle trapped in earth's gravitational field. Recent works on the gravitational well problem indicated that it can serve to shed some light on the upper bounds of various NC parameters by connecting NC theoretical results \cite{bert0, bert1, RB, ani} with the experimental data found by the GRANIT group \cite{nes1, nes2, nes3}. However, such connections are physically meaningful only if the theoretical model is constructed in such a way that preserves the space time symmetries, namely the Galilean symmetries. 

In \cite{ani} the model was constructed from a NC field theoretic platform so that the effect of time-space noncommutativity on the system can be examined. Interestingly, the singularly important result of \cite{ani} was to show that the underlying time-space sector of the NC algebra is instrumental in introducing non-trivial NC effects in the energy spectrum of the system to first order perturbative level. This lead to an upper bound estimation of the time-space NC parameter. Incedentally, apart from bringing out the effect of time-space noncommutativity on the system, the field theoretic nature of the construction also gives a perfect platform to analyse the space time symmetries. Therefore it is only natural to investigate the space time symmetries of the model discussed in \cite{ani}. In this paper we have done a thorough symmetry analysis of the same. 

Following the Noether's theorem we have worked out various transformation generators and their algebra. The spatial translation and rotation generators along with the time-evolution generator formed a closed sub-algebra of the larger Galileo group, namely the Euclidean algebra $E\left(2\right)$. The higher Galilean algebra required the construction of the boost generators. In the present non-relativistic case, the boost generators are not a part of the total angular momentum tensor, one therefore has to start from the scratch and derived them using some basic principles. Since the field variables of our theory can be interpreted as one-particle wave functions of quantum mechanics, we assumed that they can change at most by a phase factor under infinitesimal Galileo boost transformations. By demanding the covariance of the equation of motion under these Galileo boosts we derive some differential equations of the boost parameter and solving them we derived the boost generators. The presence of the external gravitational field discriminates between two boost generators, one parallel to the gravitational field and the other perpendicular to it. We found that both boost generators preserve the Galilean symmetry. Interestingly, this is in contrast to earlier results of a NC Schr\"{o}dinger theory interacting with an external electric field \cite{bcsgas}, where boost perpendicular to the direction of the external field was not a symmetry of the theory. 

The algebra among all the generators are explicitly computed and they are seen to form a closed Galilean algebra. Thus the system preserves the Galilean symmetry on the NC plane. This is a reassuring result since it shows that the comparison of the theoretical predictions of the model in \cite{ani} with the experimental results of \cite{nes1, nes2, nes3} had indeed been physically meaningful.

A worthy by-product of the analysis is found in the F--JB between the spatial translation generator $P_{x}$ and the time-evolution generator $H$, this naturally produces the gravitational force acting on the trapped particle. It shows yet another significant effect of time-space noncommutativity on the gravitational well problem. The force acting on the neutron is found to be increased by a factor $\frac{\eta \tilde{m} g}{\hbar}$. In this connection we also discuss the existing upper-bounds on different NC parameters and using them we found that presence of time-space noncommutativity can increase the gravitational pull on the trapped particle by at most $0.147 \times 10^{-6}\%$.


\end{document}